\documentclass[amsmath,amssymb,twocolumn,prl]{revtex4}
\usepackage{graphicx}

\newcommand{\beq}{\begin{equation}}
\newcommand{\eeq}{\end{equation}}
\newcommand{\beqa}{\begin{eqnarray}}
\newcommand{\eeqa}{\end{eqnarray}}

\begin{document}

\title{Viscoelasticity and metastability limit in supercooled liquids}

\author{Andrea Cavagna$^{\dagger}$, Alessandro Attanasi$^{\ddagger}$, and 
Jos{\'e} Lorenzana$^{\dagger}$}

\affiliation{
$^{\dagger}$ Centre for Statistical Mechanics and Complexity - INFM and ISC - CNR, 
Via dei Taurini 19, 00185 Roma, Italy
\\
$^{\ddagger}$ Dipartimento di Fisica, Universit\`a di Roma ``La Sapienza'',
P.le Aldo Moro 2, 00185 Roma,  Italy
}

\date{June 1, 2005}

\begin{abstract}
A supercooled liquid is said to have a kinetic spinodal if a temperature $T_{\rm sp}$ 
exists below which the liquid relaxation time exceeds the crystal nucleation time.
We revisit classical nucleation theory taking into 
account the viscoelastic response of the liquid to the formation of crystal 
nuclei and find that the kinetic spinodal is strongly influenced by elastic
effects. We introduce a dimensionless parameter 
$\lambda$, which is essentially the ratio between the infinite
frequency shear modulus and the enthalpy 
of fusion of the crystal. In systems where $\lambda$ is larger than a critical value $\lambda_c$ 
the metastability limit is totally suppressed, independently of the surface tension. 
On the other hand, if $\lambda<\lambda_c$ a kinetic spinodal is present and the time needed to 
experimentally observe it scales as $\exp[\omega/(\lambda_c-\lambda)^2]$, where $\omega$ 
is roughly the ratio between surface tension and enthalpy of fusion. 
\end{abstract}

\vglue 0.1 truecm 


\maketitle

When a liquid is cooled below its freezing point without forming a
crystal, it enters a metastable equilibrium phase known as supercooled \cite{skripov}. 
A supercooled liquid is squeezed  in an uncomfortable time region: if we are too {\it fast} in measuring its 
properties, the system cannot thermalize and an off-equilibrium {\it glass} is formed; on the other
hand, if we are too {\it slow}, the system has the time to nucleate the solid, and we obtain an
off-equilibrium {\it polycrystal}, and, eventually, a thermodynamically stable crystal 
\cite{kauzmann48,turnbull69}. 
What is the maximum degree of supercooling a metastable equilibrium liquid can reach~? 

The tricky point about this question is that it mixes aspects of the experimental protocol, 
with intrinsic properties of the system. 
If we stick to cooling the system linearly in time, there is a minimum cooling velocity 
below which the system is bound to crystallize \cite{turnbull69}. 
This velocity is inversely proportional   
to the minimum nucleation time as a function of temperature \cite{nose}: 
we cannot cool slower than this minimum cooling rate, otherwise crystallization 
occurs. On the other hand, as we cool, the relaxation time increases steeply, and therefore the system 
necessarily leaves the supercooled phase, and becomes a glass, at the temperature where the 
relaxation time becomes too large for this minimum cooling rate.

To penetrate deeper in the supercooled region, one can use an {\it ad hoc}
nonlinear cooling protocol:
cool fast close to the temperature where crystallization is a concern \cite{nose}, and
where relaxation time is still small; and slow down at lower temperatures, to cope with the 
increasing relaxation time, once the nucleation time starts raising again.
Therefore, we may think that the unique limitation to
the extent of supercooling is given by our capability of cooling slow and fast enough a sample,
and that in principle there is no bound to supercooling a metastable equilibrium liquid.

In fact, our experimental capability is {\it not} the only li\-mi\-ta\-tion to supercooling a system. 
If at a certain
tem\-pe\-ra\-ture the relaxation time of the liquid $\tau_{\rm R}$ exceeds the nucleation time of the crystal 
$\tau_{\rm N}$, no equilibrium measurements can be performed  on the liquid sample and the supercooled 
phase does not exist anymore. Such a temperature is called {\it
kinetic spinodal} $T_{\rm sp}$, and it marks 
the metastability limit of the supercooled phase \cite{kauzmann48,skripov}. 
Such a  metastability limit does not depend on the cooling protocol, and is rather an 
intrinsic property of the sample. 
Below $T_{\rm sp}$ the only equilibrium phase (either stable or metastable) is the crystal,
whereas glassy and polycrystalline off-equilibrium configurations can be obtained if we cool fast or slow
enough, respectively. 
The aim of this work is to answer the following question: what is the main mechanism determining
whether or not a metastability limit is present in a given li\-quid~? 

Our answer to the question above is that this me\-cha\-nism is {\it viscoelasticity}.
The central idea is that on time scales shorter than the structural relaxation time, deeply supercooled liquids 
exhibit a solid-like response to strains \cite{dyre}. In particular, the inclusion 
of a crystalline nucleus in the liquid matrix produces a strain \cite{zonko02,zonko03}, and
thus a long-range elastic response \cite{roi78,cah84,bus03,review}. Therefore, the thermodynamic drive
for crystal nucleation gets depressed by an elastic contribution, 
which depends on the ratio between the relaxation time $\tau_{\rm R}$ and the time 
scale $\tau_{\rm N}$ over which nucleation occurs. In this way a self-consistent equation 
involving nucleation and relaxation times is obtained. 

We will show that liquids can be classified according to the magnitude of their elastic response
relative to their bare thermodynamic drive to crystallization (that is the liquid-crystal Gibbs 
free energy density difference $\delta G$). We find that elastic effects can be large enough to
suppress the metastability limit, opening the possibility to supercool a liquid down to the point 
where its entropy reaches the entropy of the solid, the so-called Kauzmann temperature $T_{\rm k}$ \cite{kauzmann48}. 
Moreover, even if a metastability limit is present, the viscoelastic response determines how large is the relaxation 
time at the kinetic spinodal, and it thus establishes whether  the metastability limit
is accessible within the maximum experimental time, or is hidden by the off-equilibrium glassy phase.

According to classical nucleation theory (CNT), the Gibbs free energy barrier to nucleation is given by
\cite{turnbull49},
\beq
\Delta G = \frac{a\, \sigma^d}{[\delta G(T)]^{d-1}} \ ,
\eeq
where $d$ is the dimension, 
$\sigma$ is the surface tension, $\delta G$ is the (positive) Gibbs free energy difference per unit volume
between the two phases, and $a$ is a numerical constant (for a spherical
nucleus in $d=3$, $a=16\pi/3$).

When the nucleus of the stable phase forms in an elastic background, there is an extra energetic 
price that has to be paid \cite{roi78,cah84,bus03,review}.
Because of the long-range nature of the elastic strains, 
the elastic price is proportional to the nucleus volume, and thus it corrects
the bare thermodynamic drive $\delta G$,
\beq
\Delta G =  \frac{a\, \sigma^d}{\left[\delta G(T) - E_{\rm elastic}\right]  ^{d-1}} \ .
\eeq
In a perfect isotropic solid, assuming that the elastic parameters are the same in the two phases,
the elastic energy density is given by $E_{\rm elastic}=\epsilon_0\, G_\infty$
where $G_\infty$ is the (constant) elastic shear modulus, and where \cite{review},
\beq
\epsilon_0 = \frac{(d-1)\,K}{d\,K +2(d-1)\, G_\infty} \left(\frac{\delta v}{v}\right)^2 \ ,
\eeq
is a dimensionless constant: $K$ is the bulk modulus and $\delta v/v$
is the stress-free volume misfit between the 
two phases. In a viscoelastic liquid we have a stress-relaxation function, or
time-dependent shear modulus, $G(t)$ \cite{ferry}. In this case,
Schmelzer and coworkers have shown that the time-dependent elastic contribution can be written as,
\beq
E_{\rm elastic} = \epsilon_0 G_\infty f(t)
\label{ela}
\eeq
where the infinite frequency shear modulus is, $G_\infty=G(t=0)$, 
and the dimensionless function $f$ is \cite{zonko02,zonko03},
\beq
f(t) = \frac{1}{t}\int_0^t dt'\ G(t')/G_\infty \ .
\eeq
For a Maxwell liquid \cite{ferry} $G(t)=G_\infty e^{-t/\tau_{\rm R}}$ 
and thus, $f(t/\tau_{\rm R})=(1-e^{-t/\tau_{\rm R}} )  \tau_{\rm R}/t$.
We shall not assume the Maxwell form, but we shall still make the hypothesis that the
stress-relaxation function $G$ depends on the ratio $t/\tau_{\rm R}$, where $\tau_{\rm R}$ is
the liquid's structural relaxation time. In this case
also $f=f(t/\tau_{\rm R})$. In log scale the function $f$ typically has a step-like shape,
going from  $1=f(0)$ to $0=f(\infty)$ with a sharp drop at $f(1)$.
The physical meaning is clear: for times much larger than the relaxation time the liquid 
is able to relax the stress, and there is no elastic contribution, while for times 
much smaller than $\tau_{\rm R}$ it responds as a solid, with finite shear modulus.

What is the time $t$ we have to plug into $E_{\rm elastic}(t/\tau_{\rm R})$~?
The viscoelastic contribution to the nucleation barrier is due to the formation 
of a crystal nucleus, and this happens on a time scale of the order of 
the nucleation time. Therefore we must set  $t=\tau_{\rm N}$, and  write,
\beq
\tau_{\rm N}=\tau^0_{\rm N} \exp\left\{
\frac{a\, \sigma^d}{k_{\rm B} T\ \left[\delta G(T) - 
E_{\rm elastic}(\tau_{\rm N}/\tau_{\rm R})\right]^{d-1}}\right\} 
\label{lafo}
\eeq
From this formula we clearly see that the main effect of the elastic
contribution is to increase the nucleation time. This effect becomes dramatic
in solid-solid phase transitions, where nucleation may be totally suppressed, 
even with a nonzero thermodynamic drive $\delta G$, whenever $\delta G(T) < 
E_{\rm elastic}=\epsilon_0 G_\infty $ \cite{roi78,cah84,review,bus03}. In liquids, however, 
the stress is always relaxed for long enough times, and eq.(\ref{lafo}) always 
admits a solution $\tau_{\rm N}(T)$, as long as $\tau_{\rm R}<\infty$.

In  order to solve eq.~\eqref{lafo},
we have to discuss the $T$-dependence of the various quantities.
For the relaxation time we shall assume a Vogel-Fulcher-Tamman (VFT) form, which is known to describe very
well a wide variety of supercooled liquids on a large range of times \cite{angell88},
\beq
\tau_{\rm R} = \tau_{\rm R}^0 \exp\left( \frac{\Delta}{T-T_{\rm k}}\right) \ .
\label{relax}
\eeq
This form of the relaxation time implies a singularity at $T_{\rm k}$,
which is normally identified with the Adam-Gibbs thermodynamic transition  
\cite{adam65}, and it is approximately equal to the temperature where 
the liquid-crystal entropy difference $\delta S$ vanishes.
The Gibbs free energy difference is often fitted to a linear
form, $\delta G = (1-T/T_{\rm m}) \delta h/\nu$, 
where $T_{\rm m}$ is the melting temperature, $\delta h$
is the molar enthalpy of fusion, and $\nu$ is the molar volume of the crystal \cite{kiselev01,weimberg02}. 
However, recalling that $\delta G= - \partial (\delta S)/\partial T$, consistency with (\ref{relax}) 
requires that $\delta G$ must reach its maximum
at $T_{\rm k}$. Therefore, we shall write,
\beq
\delta G(T) =  g(T/T_{\rm m})  \ \left( 1-T_{\rm k}/T_{\rm m} \right) \ \delta h/\nu
\eeq
where the dimensionless function $g$ satisfies the relations: $g(1)=0$ and $g(T_{\rm k}/T_{\rm m})=1$. 
The surface tension $\sigma$ will be assumed to be a constant, which is a reasonable hypothesis 
within CNT \cite{kiselev01}.

The prefactor of nucleation $\tau_{\rm N}^0$ in (\ref{lafo}) is significantly $T$-dependent.
According to CNT, $\tau_{\rm N}^0$ is proportional to the product of 
an elementary thermal time scale, $h_{\rm plank}/k_{\rm B}T$, and an Arrhenius term due to 
the barrier $B$ a particle has to overcome to cross the nucleus interface, $\exp(B/k_{\rm B}T)$
\cite{turnbull49}. It is customary to approximate such term by $1/D$ ($D$ is the self-diffusion coefficient), 
and by using the Stokes-Einstein (SE) relation, to write 
$\tau_{\rm N}^0\sim 1/D\sim\tau_{\rm R}$ \cite{turnbull69}.
However, as the system becomes viscous, the SE relation breaks down \cite{rossler90,fujara92,cicerone96},
and typically $1/D \ll \tau_{\rm R}$. 
Indeed, in the extreme case of solids, even though the shear viscosity (and thus $\tau_{\rm R}$) is 
infinite, $D$ remains nonzero. 
Thus, the prefactor $\tau_{\rm N}^0$
remains much smaller than the relaxation time at low temperatures, 
and it cannot by itself grant
the fact that $\tau_{\rm N} \gg \tau_{\rm R}$.

A second delicate point about $\tau_{\rm N}^0$ is its dependence on
the sample's volume. The quantity studied
by CNT is not directly the nucleation time, but the
nucleation rate $J$, that is the number
of nuclei per unit time, per unit volume. $J$  is 
a constant for large enough samples and  
therefore the nucleation time of a given sample of volume $V$
scales as $1/V$ \cite{turnbull69}. Taking very small samples is indeed a well known trick to 
increase nucleation time. However, there is a limit to this procedure, given by
the volume $v_0$ below which the $1/V$ scaling breaks down: below $v_0$
the nucleation rate is not anymore constant, either because surface effects
become dominant over bulk properties, or because the sample size is smaller
than the critical nucleus \cite{paolo}.
The theoretical metastability limit of a liquid can be defined as the point where the {\it largest}
possible nucleation time is surpassed by the relaxation time, and therefore we set  
$\tau_{\rm N}= 1/(v_0 J)$. Anyhow, we recall that the
results are rather insensitive to the choice of the reference volume, since this enters
only in the prefactor $\tau_{\rm N}^0$, and not in the most relevant exponential factor.

We measure all times in equation (\ref{lafo}) in units of $\tau_{\rm N}^0$, and define 
$x= \log(\tau_{\rm N}/\tau_{\rm N}^0)$ and $y=\log(\tau_{\rm R}/\tau_{\rm N}^0))$.
In this way equation (\ref{lafo}) can be rewritten as,
\beq
T' x = \omega \ \left[\ g(T') - \lambda\, f\left(e^x/e^y \right) \ \right]^{1-d}
\label{cosimo}
\eeq
where $T'=T/T_{\rm m}$ is the reduced temperature, and,
\beqa
\omega &=& \frac{a\sigma^d}{k_{\rm B}T_{\rm m}} 
\left[ \frac{\nu}{(1-T_{\rm k}/T_{\rm m})\, \delta h\, }\right]^{d-1}
\\
\lambda &=& \frac{\epsilon_0 \nu G_{\infty}} {(1-T_{\rm k}/T_{\rm m})\, \delta h \,} \ .
\eeqa
Two more dimensionless parameters are present in the theory: they are
the rescaled Kauzmann temperature $T'_{\rm k}=T_{\rm k}/T_{\rm m}$ and
the rescaled VFT barrier $\Delta'=\Delta/T_{\rm m}$. They both enter in the relaxation time,
and thus in $y=y(T'/T'_{\rm k},\, \Delta')$.
If we set $\lambda=0$, $T_{\rm k}=0$ and $g=(1-T/T_{\rm m})$, we recover the 
classic expression for the nucleation
time, where  $\exp(\omega)$ gives the scale of the minimum nucleation time 
(see, for example, \cite{kiselev01}). 
On the other hand, for $\lambda \neq 0$ we have elastic corrections to the nucleation time.

Equation (\ref{cosimo}) can be easily studied numerically once the forms of $g$ and $f$ are specified. 
However, the key physical features can be worked out in full generality. 
First, we note that for $T\to T_{\rm m}$, $g\to 0$ whereas $y$ remains finite,
and thus the solution $x$ of (\ref{cosimo}) diverges. As expected, $\tau_{\rm N} \to \infty$ at the melting
point. In the opposite limit $T\to T_{\rm k}$ a
 graphical study of equation (\ref{cosimo}) shows that for $\lambda <1$ the nucleation
time remains finite, while $y\to\infty$, and thus a metastability limit $T_{\rm sp} > T_{\rm k}$ 
exists. 
On the other hand, 
for $\lambda>1$, $\tau_{\rm N}\to \infty$ for $T\to T_{\rm k}$: in this case both relaxation and
nucleation times diverge at $T_{\rm k}$, but they still may cross at a kinetic spinodal. 
To check this last possibility we set $x=y$ in equation (\ref{cosimo}), and see whether there is
a solution $T_{\rm sp}>T_{\rm k}$ of this equation,
\beq
T' \left( \Omega + \frac{\Delta'}{T'-T'_{\rm k}}\right) = 
\omega \left[ g(T') - \lambda \, f(1) \right]^{1-d} \ .
\label{spin}
\eeq
In this formula the factor $\Omega=\log(\tau_{\rm R}^0/\tau_{\rm N}^0)$ is not relevant, 
since it is large in modulus only at low temperatures, where however the factor 
$(T'-T'_{\rm k})^{-1}$ dominates. We recall that, by construction, $f(1)$ is a
number close to $1/2$ (in Maxwell's case $f(1)=1-1/e=0.63$). 
The l.h.s. of equation (\ref{spin}) diverges at $T'_{\rm k}$, whereas the
r.h.s. is finite at $T'_{\rm k}$ and diverges 
at a temperature larger than $T'_{\rm k}$, defined by $g(T')-\lambda f(1)=0$. 
Given that $g(T'_{\rm k})=1$, we
conclude that for $\lambda < 1/f(1)$ there is a solution $T'_{\rm sp}$, and thus a 
metastability limit, while for $\lambda > 1/f(1)$ there is no solution in the physical interval 
$ g(T') - \lambda \, f(1) \ge 0$. 
 
\begin{figure}
\includegraphics[clip,width=3.4in]{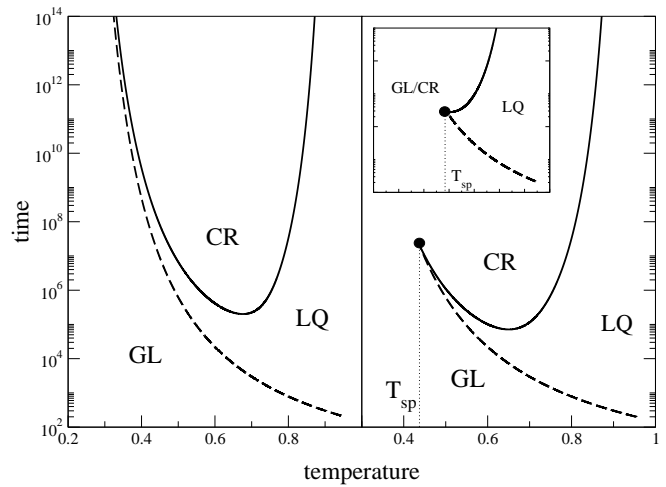}
\caption{
Nucleation and relaxation vs reduced temperature $T'=T/T_{\rm m}$, 
for different values of the parameter $\lambda$. 
Full line, nucleation: $x(T')=\log(\tau_{\rm N}/\tau_{\rm N}^0)$. Broken line, relaxation:
$y(T)=\log(\tau_{\rm R}/\tau_{\rm N}^0))$.
LQ stands for equilibrium liquid, GL for off-equilibrium glass, 
and CR for off-equilibrium polycrystal. 
Left panel: for $\lambda = 2$, larger than $\lambda_c=1.58$, there is no kinetic spinodal. 
Right panel: for $\lambda = 0.3$, smaller than $\lambda_c$, a kinetic spinodal exists at $T_{\rm sp}$. 
Inset: for $\lambda =0.03$ the difference between glass and polycrystal is particularly ill-defined. 
The other parameters are: $\omega=4,\ \Delta/T_{\rm m}=4,\ T_{\rm k}/T_{\rm m}=0.2$. For $f(t/\tau_{\rm R})$ 
we have taken a  Maxwell form, while $g(T')$ is a cubic.
}
\label{uno}
\end{figure}

We recall that $\lambda$ is, up to factors of order 
one, the ratio between the characteristic scales, $G_\infty$ and $\delta h$, of the two competing mechanisms at work: 
the elastic forces, depressing nucleation, and the thermodynamic drive to crystallization,
enhancing nucleation. What we have found is a critical value, 
\beq
\lambda_c = 1/f(1) \ ,
\eeq
separating systems with a kinetic spinodal, from systems without a kinetic spinodal.
In those liquids where $\lambda < \lambda_c$ the metastability limit may be shifted 
at lower $T$ by elasticity, but it still exists, while for $\lambda > \lambda_c$ elastic effects
are strong, and the nucleation time is amplified enough to completely suppress the kinetic spinodal.
This distinction is, of course, only valid to logarithmic accuracy: 
whenever $\lambda\sim\lambda_c$, the nucleation time 
may be only {\it slightly} larger than the relaxation time, and it is wise to assume
that the liquid is not really stable against crystallization.

In Fig.1 we show  
solutions of eq.(\ref{cosimo}) for $\lambda>\lambda_c$ (left panel), and $\lambda < \lambda_c$ (right panel).
In the first case, there is no spinodal, and thus the three phases, glass, supercooled liquid, and 
polycrystal, are well defined. In particular, the off-equilibrium glassy and polycrystalline states
are well separated from each other by the `buffer' liquid phase. When $\lambda < \lambda_c$, however,
the situation is rather different. The equilibrium liquid phase is still well defined 
as long as it exists, that is for $T>T_{\rm sp}$; however, it is hard to say what is 
the physical difference between the two off-equilibrium phases, i.e. glass and polycrystal. 
Although it may be operationally meaningful to distinguish glass from polycrystal
when $T\gg T_{\rm sp}$, there seems to be no obvious {\it qualitative} difference when $T\sim T_{\rm sp}$, or,
even worse, when $T<T_{\rm sp}$. Close to the spinodal point, whether the system falls out 
of equilibrium by crossing the broken or the full line, is irrelevant. Glass and polycrystal blend into
a single off-equilibrium phase \cite{cavagna03}.

In systems with $\lambda > \lambda_c$ 
the supercooled phase is well defined down to $T_{\rm k}$, where the relaxation time  diverges.
In this case, a solution of Kauzmann paradox (for example, in terms of a thermodynamic transition 
at $T_{\rm k}$) must be found. On the other hand, when a spinodal is present the supercooled 
liquid ceases to exist at $T_{\rm sp}$, which is Kauzmann's resolution of the paradox \cite{kauzmann48}.

When a kinetic spinodal exists, is it within the boundaries of experimental observation ? To answer this
question we have to define a spinodal time, $\tau_{\rm sp} = \tau_{\rm R}(T_{\rm sp})$, 
and compare it to the maximum time of the experiment, $\tau_{\rm exp}$.
From eq.~(\ref{spin}) we obtain (for $d=3$),
\beq
\tau_{\rm sp} \sim \exp\left[\frac{\omega}{(1-\lambda/\lambda_c)^2}\right] \ .
\label{tau}
\eeq
Therefore, as expected, the {\it position} of the kinetic spinodal depends also on the parameter
$\omega$: a large value of $\omega$ implies a large value of the surface tension $\sigma$ compared to the 
drive to crystallization $\delta h$, and thus a larger nucleation time \cite{turnbull69}. Therefore,  
when $\omega$ is significantly larger than $(1-\lambda/\lambda_c)^2$, the spinodal time may be too large 
compared to the experimental time, and thus the metastability limit be experimentally unaccessible \cite{tigi}.

Conventionally, a good glass-former is a system which does not crystallize easily under 
a linear cooling, i.e. a system whose minimum cooling rate $r_{\rm min}$ is not too large. Given that 
$r_{\rm min}\sim 1/\tau_{\rm min}$, where $\tau_{\rm min}$ is the minimum nucleation time as a function of $T$ \cite{nose},
the larger $\tau_{\rm min}$, the better the glass-former.
To what extent $\tau_{\rm min}$ depends on $\lambda$ ? When $\lambda \ll \lambda_c$ the kinetic spinodal
typically  takes place at a point where the nucleation time is still a decreasing function of $T$ (Fig.~1, inset of 
right panel). In such case  $\tau_{\rm min}$ is effectively given by the spinodal point. 
From (\ref{tau}) we therefore have $r_{\rm min}\sim \exp[-\omega/
(1-\lambda/\lambda_c)^2]$, and the glass-forming capability of the system is strongly dependent on $\lambda$. 
On the other hand, when  $\lambda \sim \lambda_c$ or $\lambda > \lambda_c$,  $\tau_{\rm min}$ 
is given as usual by the minimum of the nucleation curve. In this case $\tau_{\rm min}\sim \exp(\omega)$, 
the glass-forming properties depend almost exclusively on $\omega$, whereas their dependence on $\lambda$ is weak.

In this work we studied the effects of viscoelastic response on the 
metastability limit of supercooled liquids. We defined a
parameter $\lambda$ whose value rules whether a kinetic spinodal exists 
or not. This parameter encodes the competition between elastic response and thermodynamic 
drive to 
crystallization: for values of $\lambda$ below $\lambda_c$ elasticity is relatively 
weak, the drive to crystallization always wins, and a metastability limit is present.
For $\lambda>\lambda_c$ the elastic response is so large that nucleation in inhibited
compared to relaxation, and a metastability limit is not present.

We thank A. Angell, G. Biroli, C. Di Castro, I. Giar\-di\-na, T.S. Grigera, and P. Verrocchio for many useful 
comments and clarifying discussions.


\begin{thebibliography}{10}
\bibitem{nose}
This point is known as the ``nose'' of the the Time-Temperature-Transformation
  (TTT) curve.
\bibitem{tigi}
In this case the spinodal is hidden by the glass transition, i.e. $T_{\rm sp}<
  T_{\rm g}$, where $T_{\rm g}$ is defined as $\tau_{\rm exp}=\tau_{\rm R}(T_{\rm g})$.
\end{thebibliography}
\end{document}